\documentstyle[prl,aps,floats,twocolumn,epsfig]{revtex}

\newcommand{\beq}{\begin{equation}}
\newcommand{\eeq}{\end{equation}}
\newcommand{\bea}{\begin{eqnarray}}
\newcommand{\eea}{\end{eqnarray}}

\def\laq{~\raise 0.4ex\hbox{$<$}\kern -0.8em\lower 0.62
ex\hbox{$\sim$}~}
\def\gaq{~\raise 0.4ex\hbox{$>$}\kern -0.7em\lower 0.62
ex\hbox{$\sim$}~}

\def \ra {\rightarrow}
\def \la {\lambda}

\def \b {\beta}
\def \a {\alpha}
\def \ap {\alpha^{\prime}}

\def \ga {\gamma}

\def \da {\delta}
\def \ep {\epsilon}
\def \r {\rho}

\def \ti {\tilde}

\def \fp {\dot{\phi}}
\def \fpp {\ddot{\phi}}
\def \tH {\tilde H}
\def \tr {\tilde \rho}
\def \tp {\tilde p}
\def \tf {\tilde \phi}
\def \tV {\tilde V}

\begin{document}

\par
\begingroup
\twocolumn[%

\begin{flushright}
BA-TH/01-409\\
gr-qc/0105082\\
To appear in {\bf Phys. Rev. D}\\
\end{flushright}
\bigskip

{\large\bf\centering\ignorespaces
Dilatonic Interpretation of the Quintessence?
\vskip2.5pt}
{\dimen0=-\prevdepth \advance\dimen0 by23pt
\nointerlineskip \rm\centering
\vrule height\dimen0 width0pt\relax\ignorespaces
 M. Gasperini 
\par}
{\small\it\centering\ignorespaces
Dipartimento di Fisica, Universit\`a di Bari, 
Via G. Amendola 173, 70126 Bari, Italy\\
and Istituto Nazionale di Fisica Nucleare, Sezione di Bari, Bari,
Italy 
\par}

\par
\bgroup
\leftskip=0.10753\textwidth \rightskip\leftskip
\dimen0=-\prevdepth \advance\dimen0 by17.5pt 
\nointerlineskip
\small\vrule width 0pt height\dimen0 \relax

We discuss the possibility that ``quintessential effects", recently
displayed by large scale observations, may be
consistently described in the context of the low-energy string effective
action, and we suggest a possible approach to the problem of the
cosmic coincidence based on the link between the strength of
the dilaton couplings and the cosmological state of our Universe. 

\par\egroup
\vskip2pc]
\thispagestyle{plain}
\endgroup

In a string theory context \cite{a}, the macroscopic and large scale
gravitational interactions are described by a low-energy
effective action which necessarily contains a scalar field -- the dilaton
-- and which can be written, in general, as the action of a
non-minimally coupled and self-interacting scalar-tensor theory. 

The dilaton $\phi$ controls the strength of the gravitational coupling
and -- in superstring models of unification -- also the strength of all
other interactions, since the expectation value of the dilaton 
should fix the fundamental ratio between string and Planck mass, and
the gauge coupling constant of GUT theories \cite{b}. At the tree-level, 
\beq
g_s^2 = \exp \langle \phi \rangle \simeq (M_S/M_P)^2 \simeq
\a_{\rm GUT}.  
\label{1}
\eeq

It may be possible (and even auspicated, in certain string-inspired
cosmological scenarios \cite{c}) that in the very early past of our
Universe the dilaton was  varying very rapidly in time, and the effective
gravitational interactions were very different from those described by
the Einstein  equations. At present, however, the dilaton has to be
frozen,  to be consistent with the observed values of the coupling
constants. The string effective action should thus contain an appropriate
potential to allow the solution $\phi=\phi_0=$ const, or a regime of 
slow-enough dilaton variation. In such a regime, in addition, the dilatonic
interactions  must become weak enough and/or short-range, so that the
action may  provide a correct description of present macroscopic
gravity. 

The potential energy of a nearly constant dilaton, on the other
hand, introduces into the gravitational equations a cosmological term,
which might  lead the Universe to a phase of cosmic
repulsion and accelerated expansion. It seems thus natural to wonder
whether the dilaton could simulate consistently, in a string theory
context, the effects of the so-called ``quintessence" \cite{1}, the
phantomatic scalar field introduced to fit recent cosmological
observations \cite{2}, and to (possibly) alleviate some problems posed
by the phenomenological need for a vacuum source with
negative pressure and very small energy density, which seems to
dominate our present Universe.  If the answer would be positive,  there
would be no need to assume the existence of new exotic scalar fields, 
and/or to invent {\em ad hoc} models of scalar interactions. 

In this paper we shall assume that the dilaton is  frozen, for
some mechanism, already in the radiation era, with a potential energy
small enough to avoid disturbing the standard cosmological evolution
down to the equilibrium epoch. 
It could be argued that a frozen scalar field with, a very small
but non-vanishing potential energy, is a model of quintessence
indistinguishable from adding a ``pure" cosmological constant to the
action. This is certanly true for a minimally coupled  field, but is
not true, in general, for scalars non-minimally coupled to the geometry
and to the other matter fields, like the dilaton. 

For non-minimally coupled fields, in fact, a cosmological transition (i.e.
a change in the equation of state of the dominant cosmological
sources) tends to shift the field away from the initial
equilibrium position, and  a subsequent transition to the frozen,
potential-dominated regime is not  at all guaranteed. The dilaton, in
particular, is coupled to the trace of the matter stress tensor, and a
constant solution is  possible in the (traceless) radiation era, as we shall
see, but not in the era of matter domination. 

Thus, even if the dilaton is ``sleeping" in the radiation era, it
necessarily ``wakes up" and starts rolling down (or up) the potential
after the equilibrium time, when the Universe enters the
matter-dominated regime. However, if the dilaton mass is not too small,
the dilaton may bounce back, and approach again the freezing
position, just when its potential energy starts to become critical. It
thus becomes a non-trivial consequence of the potential and of the
``stringy" dilaton dynamics if the decelerated, matter-dominated era is
followed by a phase of potential domination and accelerated
expansion. 

The main purpose of this paper is to discuss, in the above dilatonic
scenario, the problem of the ``cosmic coincidence" \cite{CC}, which
seems to affect the so-called ``trackers solutions" \cite{4} arising in
models with power-law or exponential \cite{5} potentials, as well as in
models of quintessential inflation \cite{6}. Our effort, in particular, is
to understand whether or not such a problem may be avoided, or
relaxed, in a dilatonic scenario in which the ratio of the matter to
scalar energy density goes asymptotically to zero, and not to a
constant like in recent attempts to solve the coincidence problem
based on bulk viscosity \cite{viscosita} and on non-minimal
scalar-tensor couplings \cite{Amendola}. The scenario discussed in this
paper is also different from other, non-minimally coupled scalar models
of quintessence \cite{7}, because the dilaton potential and 
couplings  are not {\em ad hoc}, but (in principle)  prescribed by string
theory, and  because the present value of the dilaton field is not an
arbitrary parameter, but has to be determined so as to fix a realistic
set  of  GUT coupling constants, according to eq. (\ref{1}) (the
tree-level relation between the dilaton and the fundamental constants
could be non-trivially modified, however, by loop corrections
\cite{12a}). 

Let us start our discussion with the tree-level, lowest order in $\ap$,
gravi-dilaton string effective action, minimally coupled to perfect fluid
sources:
\beq
S = -\frac{1}{2\,\lambda_s^{2}}\,\int\,d^{4}x\,\sqrt{|g|}\,e^{-\phi}
\,\left[R+(\nabla_\mu\phi)^2 +V (\phi)\right]+ S_m 
\label{2}
\eeq
($\lambda_s=M_s^{-1}$ is the fundamental string length parameter).
Consider a homogeneous, isotropic and spatially flat background. By
varying the action with respect to $g_{00}, g_{ij}$ and $\phi$, and using
the dilaton equations to simplify the $g_{ij}$ equation, we get,
respectively: 
\bea
&&
\fp^2+ 6H^2-6H\fp-V=e^\phi \r,
\label{3}\\
&&
\dot H -H\fp +3 H^2 +{V'\over 2} ={1\over 2}e^\phi p, 
\label{4}\\
&&
\fp^2+ 12H^2-6H\fp-2\fpp+6\dot H +V'-V=0, 
\label{5}
\eea
where $V'=\partial V/\partial \phi$, and we have chosen the cosmic
time gauge. Note also that we have chosen units in  which $2 \la_s^2
=1$, so that $e^\phi$ represents, in string units, the effective
four-dimensional Newton constant $16 \pi G$. The combination of the
above equations leads to the usual covariant conservation of  the
energy density: 
\beq
\dot \r +3H(\r +p)=0.
\label{6}
\eeq

It is useful, at this point, to rewrite the dilaton equation by eliminating
$H^2$ and $\dot H$. This gives the condition
\beq
\fpp+3H\fp -\fp^2 +{1\over 2} e^\phi (\r-3p) +V'+V=0.
\label{7}
\eeq
A stable solution $\phi=\phi_0=$ const is thus possible only if 
\beq
3p-\r =2 e^{-\phi}(V+V')= {\rm const},
\label{8}
\eeq
which, combined with eq. (\ref{6}), leaves only three possibilities: 
{\em i)} vacuum, $\r=p=0$, $V+V'=0$; {\em ii)}  cosmological constant,
$\r=-p=\r_0=$ const, $V+V'=-2e^{\phi_0}\r_0=$ const; {\em iii)}  
radiation, $\r=3p$,  $V+V'=0$. 

The first two cases corresponds to an accelerated de Sitter solution,
$H^2=$ constant. Only in the third case we can obtain a decelerated
background, which coincides with the standard radiation-dominated
solution provided $V(\phi_0), V'(\phi_0) \ll e^{\phi_0 }\r$. Let us thus 
suppose that, for some mechanism (to be discussed elsewhere), the
dilaton is attracted to the equilibrium position $V+V'=0$  early enough in
the radiation era, and that the potential energy is always subdominant
during the whole radiation epoch, $V(\phi_0)\ll H_{\rm eq}^2$, in such a
way as to avoid any conflict with standard big bang nucleosynthesis
(unlike other models of extended quintessence, see \cite{Steigman}). 

It should be noted that such a dilaton configuration
extremizes the effective (canonical) potential when transformed to the
Einstein frame. In the (tilded) Einstein frame variables, defined by the
conformal transformation
\beq
\ti g_{\mu\nu} =g_{\mu\nu} e^{-\phi}, ~~~~~~~~~~~
\ti \phi= \phi,
\label{9}
\eeq
the dilaton in fact is minimally coupled to the metric, but
non-minimally coupled to the fluid sources, and the
cosmological equations (\ref{3}-\ref{5}) become (in units $16\pi G=1$): 
\bea
&&
6 \tH^2 = \tr +\tV +{\dot {\phi}^2\over 2},
\label{10}\\
&&
4 \dot {\tH} + 6 \tH^2 =-\tp +\tV -{\dot {\phi}^2 \over 2},
\label{11}\\
&&
\ddot {\phi}+3\tH \dot {\phi} +{1\over 2} (\tr-3\tp) +{\tV}'=0,
\label{12}
\eea
where $\tV = e^\phi V$, $\tr = e^{2\phi} \r$, $\tp = e^{2\phi}p$, and  
the dot denotes the derivative with respect to the Einstein cosmic time,
$d\ti t= dt e^{-\phi/2}$. Thus, in the radiation era, ${\tV}'=0$ is the
necessary condition for $\phi=$ const.

Let us now approach the problem of the cosmic coincidence considering
the Einstein frame equations (\ref{10}--\ref{12}), which we rewrite in
the matter dominated era $\r=\r_m$,  $p=0$ (omitting the tilde, for
simplicity), and taking into account the possibility of a more general
(loop-induced, see below) matter-dilaton coupling, parametrized by the
function $\alpha(\phi)$:
\bea
&&
6 H^2 = \r_m +V +{\dot {\phi}^2\over 2},
\label{13}\\
&&
4 \dot {H} + 6 H^2 =V -{\dot {\phi}^2 \over 2},
\label{14}\\
&&
\ddot {\phi}+3H \dot {\phi} +{1\over 2} \alpha(\phi)\r_m +{V}'=0 
\label{15}
\eea
($\alpha=1$ corresponds to the previous, lowest-order action). 
Their combination gives 
\beq
\dot {\r_m} + 3 H \r_m  - {1 \over 2}\alpha(\phi)\r_m\dot {\phi} =0. 
\label{16}
\eeq
Assuming that the matter-dominated era starts at the equilibrium time
with $\phi=\phi_0$, $ \dot \phi=0= V'$, $V_0 \equiv V(\phi_0) \ll 6
H^2_{\rm eq}$, $\r_m \simeq 6H^2_{\rm eq}$, we ask the question: is it
possible for the phase of matter domination to evolve into a
subsequent phase dominated by the potential energy of the scalar
field, and corresponding to an effective negative pressure
$p_\phi/\r_\phi \equiv (\dot \phi^2/2 -V)/
(\dot \phi^2/2 +V)<0$ ? 

In the absence of the non-minimal coupling to matter (i.e $\a =0$, pure
general relativity) the answer would be yes, for all values of
$V_0<6H^2_{\rm eq}$. In that case, indeed, the dilaton wold keep
constant at the minimum, and the Universe would enter the
potential-dominated phase as soon as $\r_m \laq V_0$, rapidly
approaching a final regime with $p_\phi/\r_\phi =-1$, and $ H^2 =
V_0/6=$ const. However, given the (semi-)infinite range of allowed 
values for $V_0$, why $V_0 \sim \r_m (t_0) \sim H_0^2$ (as suggested
by osservations), i.e. why the potential energy of the scalar field is
just of the same order as the present matter energy density? (cosmic
``coincidence"). 

The dilaton, however, {\em is} non-minimally coupled to macroscopic
matter ($\a \not= 0$, in general), and cannot stay frozen at the 
minimum when the Universe is driven by $\r_m$ (see eq. (\ref{15})). The
above question concerning the possible advent of the quintessential
regime thus becomes:  for which values of the initial potential $V_0 <
H^2_{\rm eq}$ the dilaton may come back to the minimum, and the
Universe may enter the potential-dominated phase, with total negative
pressure? If the answer would point, only and precisely, at the value
$V_0 \sim H_0^2$, the coincidence problem would disappear. If the
answer would indicate instead a restricted range of values for $V_0$,
the problem would be nevertheless alleviated. 

The answer to the above question obviously depends on the shape of
the potential, and on the coupling $\a(\phi)$. Let us start assuming, 
first of all, that $\a$ is a constant, or that its dilaton dependence (in
the range of interest of our problem) is so weak to be consistently
neglected. For the dilaton potential, we can ask the assistance of string
theory. Although the full and detailed form of $V(\phi)$ is largely
unknown (mainly because of non-perturbative effects), we know, 
however, that in the weak coupling regime ($\phi \ra -\infty$, $g_s \ll
1$), the supersymmetry breaking potential \cite{Witten} of critical
superstring theory has to be strongly suppressed in an istantonic way, 
i.e. $V \sim  \exp(-1/g_s^2)\equiv \exp[-\exp(-\phi)]$. In the regime of
moderately strong coupling ($-\phi \sim 1$), on the other hand, the
potential must exhibit some structure, leading to a minimum $\phi_0$
such that \cite{b} 
\beq
g_s^2 =\exp (\phi_0)\simeq 0.1 - 0.01,
\label{17}
\eeq
(according to eq. (\ref{1})). Finally, in the strong coupling regime ($g_s
\sim 1$), we may expect an exponential growth of the potential, due to
the factor $ e^\phi$ induced by the transformation from the string to the
Einstein frame. Such an exponential behaviour could be suppressed, at
very strong coupling $g_s \gg 1$, by some  non-perturbative effect or
by loop corrections, but we are not concerned with this possibility here
since the matter coupling  is expected to shift the dilaton towards
negative values, and then towards the weak coupling regime (see eq.
(\ref{15})).

What is imporant, in this context, is that the potential barrier
separating the minimum $\phi_0$ from the perturbaive regime is not
infinitely high, and it is possible for the shifted dilaton to escape to
$-\infty$, provided its mass is light enough, and/or its coupling to
matter (i.e., the initial acceleration) is sufficiently strong. The possible
transition to a phase dominated by the potential energy $V_0$ thus
becomes a precise problem of balance between the initial force aiming
at moving the dilaton away from the minimum, $-\a \r /2<0$, and the
restoring force generated by the potential, $-V' >0$. 

\begin{figure}[t]
\begin{center}
\includegraphics[width=50mm]{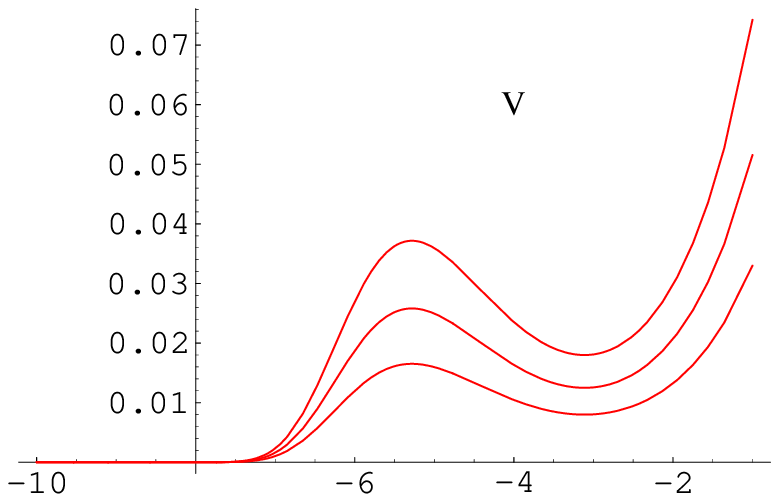}
\includegraphics[width=50mm]{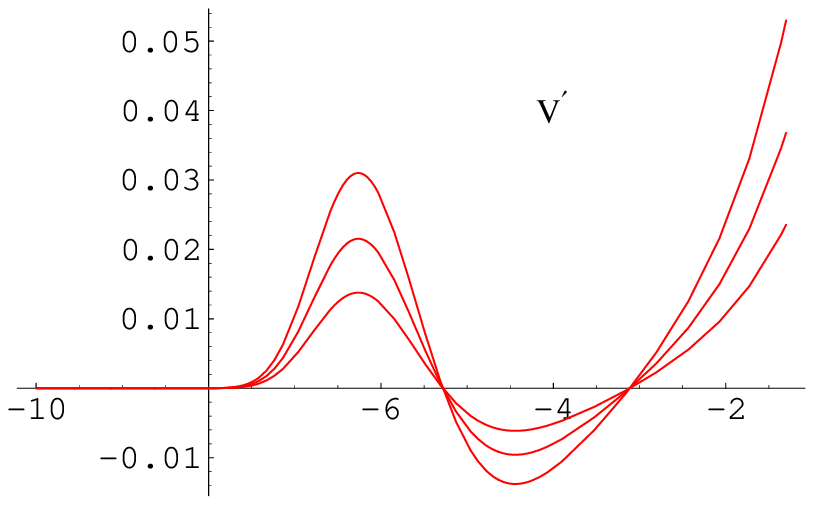}
  \end{center}
\vskip 5mm
\caption{\sl Plot of $V(\phi)$ and $V'(\phi)$ from eq. (\ref{18}), with 
$k_1=k_2=\b=\ga=1$, $ \ep=0.1$, $\phi_1=-3$. The three curves, from
top to bottom, correspond respectively to $m=1/10$, $m=1/12$,
$m=1/15$, in units of $H_{\rm eq}$.}    
\end{figure}

In order to illustrate this mechanism, let us consider  a ``minimal"
example of dilaton potential satisfying the above-mentioned string
theory requisites, and controlled only by one dimensional parameter
$m$, related to the effective (low-energy) mass of the dilaton. Such a
potential can be simply parametrized (in the Einstein frame) as follows: 
\beq
V= m^2 \left[e^{k_1(\phi_-\phi_1)}+ \beta
e^{-k_2(\phi_-\phi_1)}\right] e^{-\ep \exp\left[-\gamma
(\phi_-\phi_1)\right]}
\label{18}
\eeq
(see also \cite{Olive}),  
where $k_1,k_2, \phi_1,\ep ,\b,\ga$ are dimensionless numbers of
order one, whose precise values are not crucial for the purpose of this
paper, provided they determine a minimum around a value $\phi_0$
consistent with a superstring unification scenario (eq. (\ref{1})). For
our illustrative purpose we will choose the particular values
$k_1=k_2=\b=\ga=1$, $ \ep=0.1$, $\phi_1=-3$, in such a way that the
minimum is at $\phi_0 =-3.112...$, and $ g_s^2 =
\exp (\phi_0)\simeq 0.045$, in agreement with eq. (\ref{17}). (Many
other choices are possible, of course. In particular, the qualitative
behaviour of the potential remains the same if we increase $\ep$,
provided the values of $k_1$ and $k_2$ are also simultaneously 
increased, or the value of $\gamma$ is decreased).

\begin{figure}[t]
\begin{center}
\includegraphics[width=50mm]{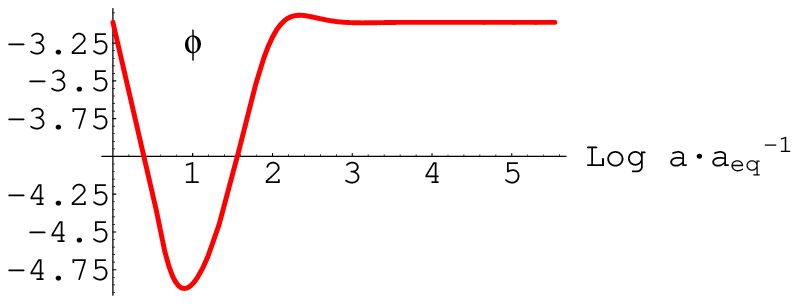}
  \includegraphics[width=50mm]{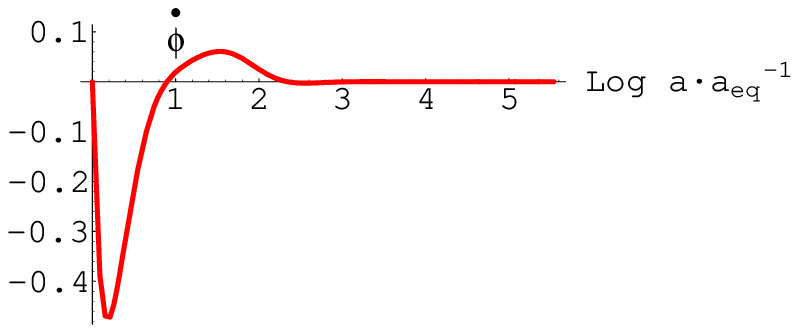}
  \includegraphics[width=50mm]{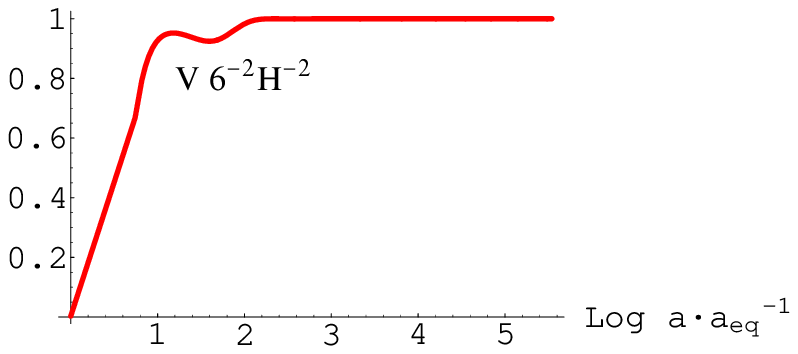}
  \includegraphics[width=50mm]{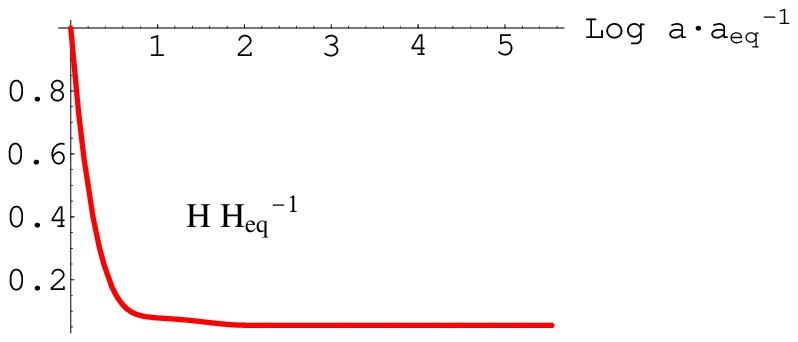}
  \end{center}
\vskip 5mm
\caption{\sl Numerical integration of eqs. (\ref{13}-\ref{16}) with
$\a=1$, $m=10^{-1} H_{\rm eq}$, and initial conditions
$\phi=\phi_0=-3.112$, $\dot\phi=0$, $a=a_{\rm eq}=1$, $H=H_{\rm
eq}=1$, $\r_m=6H^2_{\rm eq}-V_0$.}    
\end{figure}

With the above choice of parameters, the dilaton potential (\ref{18})
and its gradient $V'$ are plotted in Fig. 1, for various values of $m$. We
may note that, as $m$ is
decreased, the effective potential barrier is lowered,  the restoring
force $-V'$ becomes weaker, and it becomes easier for the dilaton to
escape from the minimum, as previously anticipated.

With such a potential, we have numerically integrated eqs.
(\ref{13}-\ref{16}), at fixed $\a$ (in particular, $\a=1$), and initial
condition at the equilibrium epoch: $\phi=\phi_0=-3.112$, $\dot\phi=0$,
$H=H_{\rm eq}$, $\r_m=6 H_{\rm eq}^2-V_0$, $a=a_{\rm eq}$. We have
found that, when the mass parameter $m<H_{\rm eq}$ is not too small
(for instance, $m=10^{-1}H_{\rm eq}$), the cosmological evolution is
typically the one illustrated in Fig. 2: the dilaton is shifted from the
minimum but is subsequently re-attracted to it, its velocity is damped
and goes to zero after some oscillations, the potential energy
asymptotically becomes critical, and the Hubble parameter freezes at
a constant value determined by $V_0$ (and corresponding to a de
Sitter equation of state $p_\phi/\r_\phi=-1$).

When the mass parameter $m$ is too small, however,
the evolution is qualitatively different, and is illustrated in Fig. 3,
where we present the results of a numerical integration with the
same initial conditions, but a mass ten times smaller, $m=10^{-2}H_{\rm
eq}$: the velocity $\dot \phi$ is still damped and goes to zero,
asymptotically, but it keeps negative for ever, and the dilaton runs
monotonically towards the perturbative regime $\phi \ra -\infty$, $V
\ra 0$.

We observe now that, in a realistic model of quintessence, the potential
energy of the dilaton should represent an important fraction of the
critical energy density (in particular \cite{2}, $V_0\simeq (2/3)\r_c
\simeq 4 H^2$) just at the present epoch, i.e. when the radiation
energy density $\r_r$ (which evolves independently from the dilaton)
has been reduced by a factor $(\r_m/\r_r)_0 =(a_0/a_{\rm eq}) \sim
10^4$ with respect to the matter energy density. This requires $V_0
\sim H_0^2$, where $H_0$ is the present value of the Hubble radius,
and since (from eq. (\ref{18})) $V_0=2m^2 \cosh (0.112)\exp[-0.1
\exp(0.112)] \simeq 2 m^2$, this implies $m \sim H_0 \sim 10^{-6}
H_{\rm eq}$. The qustion is now whether or not, for such a value of
$m$, the solution may become asymptotically dominated by the
potential. 

\begin{figure}[t]
\begin{center}
\includegraphics[width=50mm]{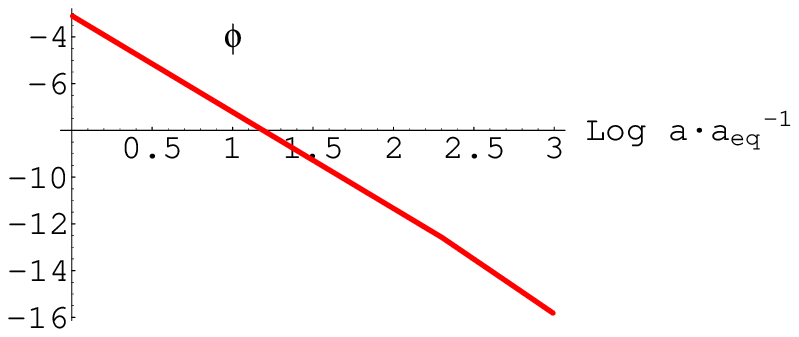}
  \includegraphics[width=50mm]{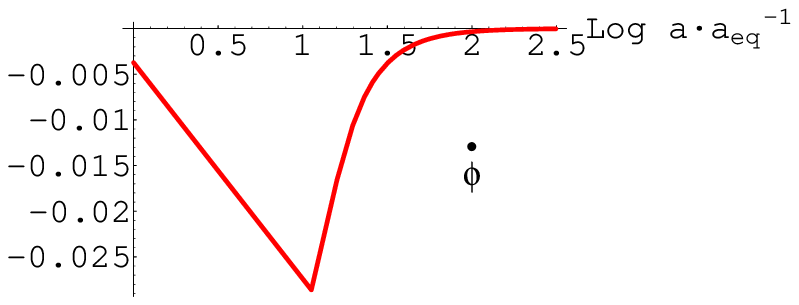}
\includegraphics[width=50mm]{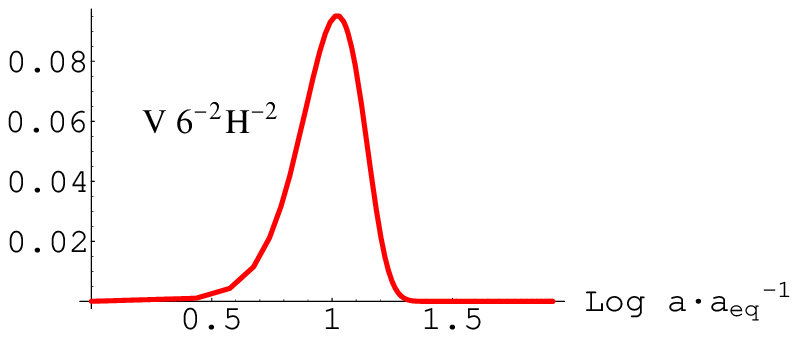}
\includegraphics[width=50mm]{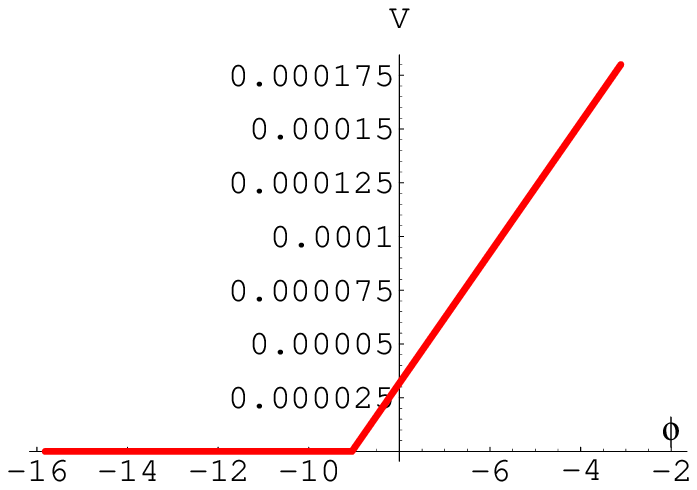}
  \end{center}
\vskip 5mm
\caption{\sl Numerical integration of eqs. (\ref{13}-\ref{16}) with
$\a=1$, $m=10^{-2} H_{\rm eq}$, and the same initial conditions
as in Fig. 2.}   
\end{figure}

The answer is strongly dependent on the value of the dilaton coupling
$\a(\phi)$. If $\a=1$, for instance, the value $m \sim H_0 \sim 10^{-6}
H_{\rm eq}$ is certainly to be excluded because, as shown by a
numerical integration, the dilaton returs to the minimum of the
potential only if $m \gaq 0.077 H_{\rm eq}$. In that case, the regime of
potential domination occurs too early, as illustrated in Fig. 2 where
$V_0\sim \r_m$  for $a\sim 10 a_{\rm eq}$, i.e. for $\r_r \sim
10^{-1}\r_m$. If $\a =10^{-1}$, instead, we find that all the values of
$m$ from $H_{\rm eq}$ down to $10^{-7} H_{\rm eq}$ are compatible
with a dilaton trapped around the minimum. 
The realistic value $m \sim 10^{-6} H_{\rm eq}$ is still to be fixed by
hand, but the choice is restricted within a limited range and,
in this sense, the coincidence problem seems to be alleviated.

It must be stressed, at this point, that the dilaton potential in the
Einstein frame also determines the mass $\tilde m$ of the (canonically
normalized) dilaton field, through its expansion around the minimum:
\bea
&&
\tilde m^2 = V''(\phi_0)= 2 m^2 \cosh(\phi_0-\phi_1)
e^{-{1\over 10}\exp(\phi_1-\phi_0)}\nonumber \\
&&
\times 
\left[1-0.1 e^{-(\phi_0-\phi_1)}-10^{-2} e^{-2(\phi_0-\phi_1)}\right] 
\simeq 2 m^2,
\label {19}
\eea
where $(\phi_0-\phi_1)\simeq -0.112$. A realistic scenario, with $m
\sim H_0 \sim 10^{-33}$ eV, thus correspond to a very long (infinite, in
practice) range for the dilatonic interactions. It follows that the
present value of the dilaton coupling, $\a_0\equiv \a [\phi(t_0)]$, has
to be strongly suppressed, to be compatible with the present tests of
the gravitational interaction. 
In particular \cite{10}, $\a_0 \laq 10^{-4}$ for
composition-dependent dilatonic couplings, strongly constrained by the
precise tests of the equivalence principle; $\a_0 \laq 10^{-2}$ for
universal dilatonic couplings, constrained by tests of post-Newtonian
gravity. 
Such phenomenological constraints can be accounted for, in principle,
by including in the string effective action the quantum loop corrections,
to all orders \cite{10} (indeed, 
they are to be included when the string coupling $g_s$ is not
negligibly small, like in the case we are considering;
higher-curvature $\ap$ corrections, on the contrary, can be safely
neglected, as the curvature scales we are considering are always small
in string units,  $\la_s^2 H^2 \ll 1$). 

The loop corrections modify the dilaton coupling to the matter sources,
and can be parametrized, in our case, by three effective dilaton ``form
factors", $Z_\r(\phi)$, $Z_p(\phi)$, 
$Z_\phi(\phi)$, appearing in the string frame action (\ref{2}), and 
defined by:
\beq
{\da S_m\over \da g_{00}}= {1\over 2} Z_\r T_{00}, ~~
{\da S_m\over \da g_{ij}}= {1\over 2}Z_p T_{ij}, ~~
{\da S_m\over \da \phi}= Z_\phi T, 
\label{20}
\eeq
($T$ is the trace of the stress tensor). The transformation to the
Einstein frame leads then to cosmological equations in which the
canonically rescaled fields $\tr, \tp, \tf$, are minimally coupled to the
metric, according to eqs. (\ref{10}, \ref{11}). The dilaton, however,
turns out to be non-minimally coupled to the trace of the fluid stress
tensor,  as in eqs. (\ref{15},\ref{16}), 
through the coupling function $\a (\phi)$ that depends
on $Z_\phi$.

In the absence of a closed and explicit expression for the loops
corrections one might assume, in the spirit of \cite{10}, that the
present value of the coupling is small because the dilaton is attracted
towards a local extremum $\phi_m$ of the coupling function. One can
then expand (to first order) $\a(\phi)=k(\phi-\phi_m)$, where $k$ is a
dimensionless number of order one.  In that case, if $\phi_m$ exactly
coincides with $\phi_0$, and with the initial position of the dilaton,
then the dilaton is decoupled from matter already in the radiation era,
it cannot be shifted from the minimum, and the coincidence problem
remains. 

If, however, the initial dilaton $\phi_{\rm eq}$ at equilibrium is slightly
shifted from the extremum, because $\phi_m\not=\phi_0$ and/or 
$\phi_{\rm eq}\not=\phi_0$, a difference $|\phi_{\rm eq}-\phi_m|\laq
10^{-1}$ is already enough (as previously mentioned) to enlarge the
allowed values of $m$ to the range $H_{\rm eq} \ra 10^{-7}H_{\rm eq}$,
which includes $m \sim H_0$. In such a case one can easily match
present observations, concerning both the cosmic equation of state and
the present  coupling of matter to scalar, long-range forces, as
illustrated in Fig. 4 for an initial dilaton slightly tilted from the
minimum. By choosing the appropriate value of $V_0$, 
the quintessential regime with critical potential and frozen Hubble
radius may start indeed  around a red-shift $a \sim 10^4 a_{\rm
eq}$, and just when $H\sim H_0 \sim 10^{-6} H_{\rm eq}$. Note also that,
in the example of Fig. 4, $\dot \phi/H$ and $\a (\phi)$  tend
asymptotically to zero,  and remain small enough to satisfy the present
phenomenological constraints \cite{10}. 

\begin{figure}[t]
\begin{center}
\includegraphics[width=50mm]{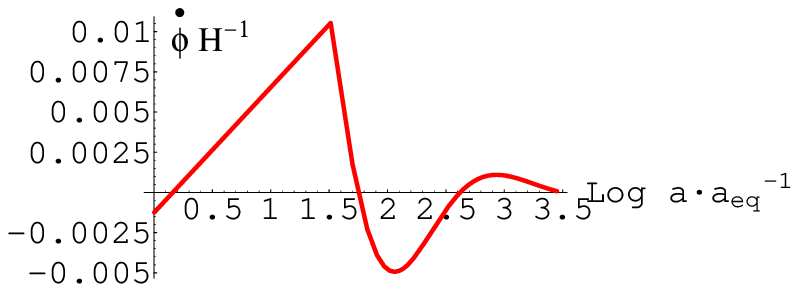}
  \includegraphics[width=50mm]{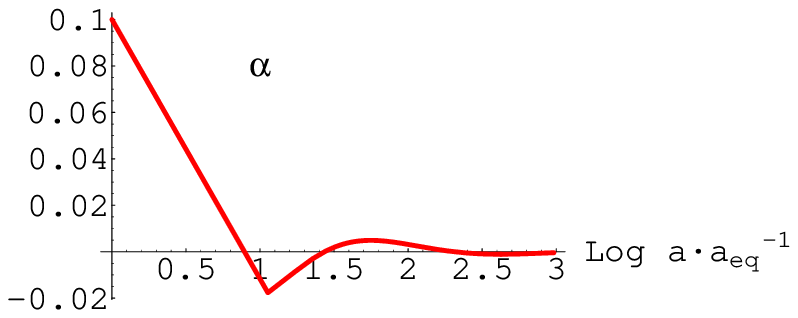}
  \includegraphics[width=50mm]{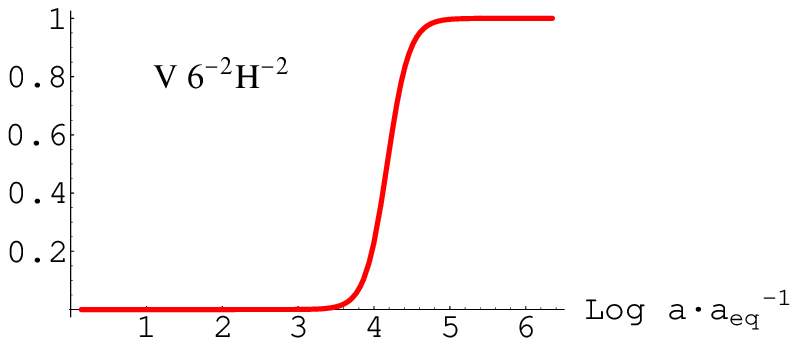}
  \includegraphics[width=50mm]{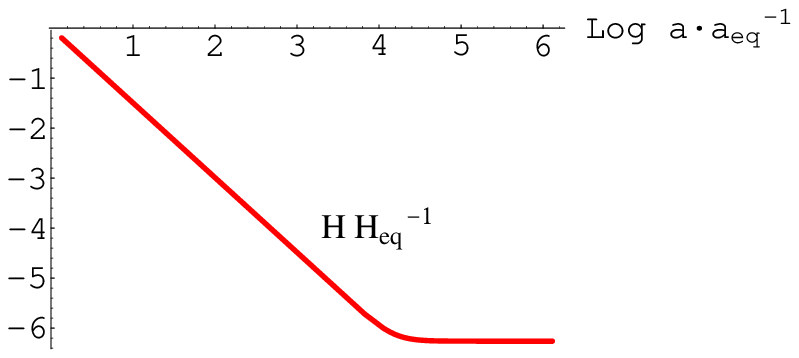}
  \end{center}
\caption{\sl Numerical integration of eqs. (\ref{13}-\ref{16}) with
 $m=10^{-6} H_{\rm eq}$, 
$\a(\phi)=k(\phi-\phi_m)$, $k=1$, $\phi_m=\phi_0$, and an initial
dilaton slightly displaced from the minimum, $\phi_{\rm
eq}=\phi_0+0.1$. The other initial conditions are the same  as in Fig. 2.}   
\end{figure}

The main conclusion of this discussion -- quite irrespective of the given
particular examples, and of their possible relevance for a complete and
fully realistic scenario -- is that a string cosmology interpretation of
quintessence suggests a close
relationship between the problem of the cosmic coincidence, and the
problem of fixing the present value of the dilaton coupling to
 matter (see also \cite{Fujii}, for previous discussions of the link
between quintessence and possible deviations from Newtonian
gravity). In particular, a strong value of the coupling 
$\a(\phi) \sim 1$ -- which is excluded by the gravitational
phenomenology -- would be {\em indirectly forbidden}, in such a
context, {\em also by the present large scale configuration} of our
Universe. Vice-versa, given $\a(\phi)$ in agreement with
phenomenology, the allowed range of the dilaton potential turns out to
be fixed, and the coincidence problem possibly alleviated, once one
assumes $V_0 <H^2_{\rm eq}$.

An important open problem, in this approach, remains however to
explain how the dilaton is attracted to the minimum of the
potential in the radiation era, and why  the vacuum energy is relaxed to a
very small value -- for instance, according to the mechanism dicussed
in \cite{Rub}. This problem requires a discussion of the initial conditions
characterizing the (post-inflationary) cosmological configuration of our
Universe, after the re-heating (and possibly pre-heating) phase, and will
be addressed in a future paper.

\acknowledgments
I wish to thank Gabriele Veneziano for many helpful discussions.  I am
also grateful to Ruth Durrer and Kerstin Kunze for interesting
conversations during the early stages of this work.

\end{document}